# Magnon-phonon coupling unmasked: a direct measurement of magnon temperature

Milan Agrawal[1], Vitaliy I. Vasyuchka[1], Alexander A. Serga[1], Alexy D. Karenowska[2], Gennadiy A. Melkov[3], and Burkard Hillebrands[1]

[1]Fachbereich Physik and Forschungszentrum OPTIMAS, Technische Universität Kaiserslautern, Kaiserslautern, 67663, Germany.

[2]Department of Physics, Clarendon Laboratory, University of Oxford, Oxford OX1 3PU, UK

[3]Faculty of Radiophysics, Taras Shevchenko National University of Kyiv, 03127 Kyiv, Ukraine

**Thermoelectric phenomena in magnetic materials present tantalizing possibilities for manipulating spin-information using heat in future 'spin caloritronic' devices[1-9]. Key to unraveling their underlying physics is to understand spin-lattice interactions, i.e. the coupling between magnons (the quanta of magnetization excitations) and phonons (the quanta of lattice vibrations). Here, we present the first measurements of the spatial distribution of magnon temperature in a magnetic system subject to a lateral thermal (i.e. phonon temperature) gradient and demonstrate that, contrary to currently accepted theory[10-12], the magnon and phonon temperatures do not differ. This result has profound implications. In particular, it re-opens the question of how the spin Seebeck effect—which allows spin currents to be produced from thermal gradients, and is arguably the most intriguing and technologically relevant thermoelectric phenomenon of all—can exist, and which physics underpins it. Specifically, it reveals that if the general framework of the current theory of the effect holds, we must adopt a new concept of spectrally non-uniform magnon temperature.**

Current understanding of a range of thermoelectric magnetic effects including the spin Seebeck effect hinges on the assumption that the coupling between the magnon and phonon systems in the materials in which they have been observed is sufficiently weak that they can establish dynamic thermal equilibria at distinct temperatures. In such a case, when a thermal (phonon) gradient $\nabla T$ is applied along a sample, it will produce a magnon density gradient in the same direction. Motion of magnons down this gradient leads to the magnon temperature $T_m$ being lifted above the phonon temperature $T_p$ in the cooler regions[10,11], and pulled below it in the hotter ones (Fig. 1). However, until now no experiment has been demonstrated capable of directly measuring and comparing the magnon and phonon temperatures in a magnetic system and thus establishing whether or not this mechanism is indeed the true one.

Magnons are bosonic quasiparticles whose population at temperature $T_m$ is described by Bose-Einstein statistics and given by $[\exp(\hbar\omega/k_B T_m) - 1]^{-1}$, where $k_B$ is the Boltzmann constant, and $\omega$ is the frequency of magnons[13]. Since each thermal magnon reduces the total magnetization of a magnetic system by one Bohr magneton, the local magnetization is a measure of the local magnon population, and hence the magnon temperature. It follows that spatial variations in the magnon temperature of a magnetic system can be determined through spatially resolved measurements of its magnetization. However, the changes in magnetization which must be detected are extremely small, making the measurement task a challenging one. The technique of Brillouin light scattering (BLS) spectroscopy offers an elegant solution. We have developed a method to study thermally induced magnetization gradients via BLS measurements of the frequency of a strongly spatially localized thermal magnon mode of known wavenumber.

Our experiments were performed using a monocrystalline (⟨111⟩) yttrium iron garnet (YIG, $Y_3Fe_5O_{12}$) film (3 mm × 10 mm, thickness 6.7 μm) epitaxially grown on a 0.5 mm thick gallium gadolinium garnet (GGG, $Gd_3Ga_5O_{12}$) substrate and magnetized by an in-plane magnetic field $B$ = 250 mT. A schematic diagram of the experimental setup is shown in Fig. 2(a). The phonon thermal gradient along the film, created by two Peltier elements, was measured using an infra-red (IR) camera with a temperature resolution of 0.1 K and a spatial resolution of 40 μm. A small laser power (7 mW) was used for the BLS measurements so as to minimize local heating and the formation of thermal gradients around the laser spot. During the measurement process the laser spot temperature was kept stable to within ± 0.3 K.

Typically, in YIG films having thicknesses of order microns, a broad spectrum of magnons is thermally excited at room temperature[13,14]. Different magnon modes can be probed using BLS technique by changing the direction of the incident-photon wavevector (i.e. the angle of the optical probing beam) relative to the magnetization of the film[15]. Owing to the weak spin-orbit coupling in YIG, the efficiency of inelastic photon scattering, and therefore the intensity of BLS signals for a given thermal magnon density, is small. However, a particular thermal magnon mode $k_{m0}$ always exists which scatters the light with higher efficiency. This mode travels along the probing light inside the film and satisfies the momentum conservation condition $|k_{m0}| = 2\eta|k_l|$, where $\eta$ (= 2.36 for YIG[16]) is the refractive index of the film and $k_l$ is the photon wavevector ($|k_l| = 1.18 \times 10^5$ rad cm$^{-1}$ for the used laser light of wavelength 532 nm). Conservation of momentum implies that photons scattered by thermal magnons of wavevector $k = k_{m0}$ propagate 'back' along their original path, in what follows we shall refer to this mode as the back-scattering magnon (BSM) mode. The position of the BSM mode is always well-defined in the magnon spectrum and under the conditions of our experiments, lies in the region of exchange-dominated magnons $|k_{m0}| = k_{m0} = 5.67 \times 10^5$ rad cm$^{-1}$ (wavelength 110 nm).

Figure 3(a) shows the position of the BSM mode on the calculated magnon dispersion curves corresponding to an in-plane magnetized YIG film at 300 K (upper curve) and 400 K (lower curve)[17,18]. Owing to its short wavelength, good coupling to the laser probing beam, and extreme sensitivity to its local magnetic environment, the BSM mode provides an excellent means to measure the spatial dependence of the magnetization—and hence magnon temperature—of a magnetic system by measuring its frequency.

In a first reference experiment, BLS measurements were carried out on a uniformly heated YIG film (i.e. without any thermal gradient) so as to measure the temperature dependence of the BSM mode. The measurements show a monotonic decline in the magnon frequency as the temperature of the YIG film increases as a result of the rising thermal magnon population (see Fig. 3(b)). Generally, in magnetic systems where the phonon temperature $T_p$ is uniform throughout, magnons eventually establish equilibrium with phonons via magnon-phonon interactions, so that the temperatures of both sub-systems are equal[10]. Accordingly, by fitting the $k_{m0}$ mode frequency dependence on $T_p$ with a third order polynomial, as shown in Fig. 3(b), the magnon frequency can be expressed as a function of the magnon temperature $T_m$.

Once the reference experiments were complete, a thermal gradient $\nabla T$ was created and maintained along the YIG strip (Fig. 2(b)). The phonon temperature profile was observed to be almost linear between the hot and the cold edges of the heat reservoirs ($\Delta T_p = 85$ K). The frequency of the BSM mode was measured simultaneously with the phonon temperature along the film parallel to the thermal gradient. In Fig. 4, the phonon temperature, $T_p$ and the magnon temperature, $T_m$, calculated from the frequency of the BSM mode, are plotted as a function of position along the YIG film. It is evident that $T_m$ follows the trend of $T_p$ within the limit of experimental error. The difference between the two temperatures along the film is shown in the inset with a 95% confidence level. The maximum difference between $T_p$ and $T_m$ is only about 2.8% of $\Delta T_p$ and, in contrast to theoretical expectations[10], this difference does not change monotonically between the hot and the cold edges.

The magnon and phonon temperatures $T_m$ and $T_p$ are coupled through the magnon-phonon interaction and, within the framework of a one-dimensional model as described in ref. [10], obey:

$$\frac{d^2T_m(x)}{dx^2}+\frac{1}{\lambda^2}[T_p(x)-T_m(x)]=0, \tag{1}$$

where $\lambda$ is a characteristic lengthscale proportional to the square root of the magnon-phonon relaxation time. Equation (1) implies that in a magnetic system where the relaxation time, and hence $\lambda$, is large, the difference in $T_p(x)$ and $T_m(x)$ will be pronounced at the boundaries. However, until the work reported here, no experiment had been demonstrated capable of directly measuring the value of $\lambda$. Moreover, theoretical estimates for $\lambda$ in YIG have varied significantly from 0.85 mm to 8.5 mm[11].

To experimentally determine the value of $\lambda$ the measured phonon temperature $T_p(x)$ was fitted with a Boltzmann sigmoid function (Fig. 4) and substituted into equation (1) to obtain a second order differential equation in $T_m(x)$ alone. This equation was solved numerically for different values of $\lambda$. In our calculations, $dT_m(x)/dx$ was assumed to be zero at the sample boundaries (since, as the magnons are confined to the sample, heat can dissipate only through phonons at the edges). From our measurements we determined the maximum possible difference between $T_m(x)$ and $T_p(x)$ at the positions $x = \pm 2$ mm, and numerically calculated the value of $\lambda_{max} \approx 0.47$ mm. This value is roughly one order of magnitude smaller than that estimated by Xiao *et al.* [11] for YIG using the experimental spin Seebeck data of ref. [2].

Our findings have significant implications for the spin Seebeck effect. Current thinking attributes the spin Seebeck effect to the existence of a temperature difference between the magnon and phonon baths in a magnetic system subject to a lateral thermal gradient. The fact that our measurements indicate no such difference therefore presents a puzzle. However, a potential explanation presents itself: Our findings rule out the possibility that the temperature of short-wavelength thermal magnons ($k_m \geq 10^6$ rad cm$^{-1}$) that strongly dominate the room-temperature magnetization differs from the phonon temperature $T_p$. Indeed, for such magnons spin-lattice thermalization is dominated by three-particle (Cherenkov) processes[13,19]. The probability of these processes, which involve the creation or annihilation of a phonon by a magnon but do not change the total magnon population, is proportional to $k^2_m$. Accordingly, the relaxation channel they provide reduces rapidly with reducing $k_m$. In light of this, it is plausible that the magnon temperature is *wavenumber dependent* (i.e. $T_m = T_m(k_m)$) and that a difference between $T_m$ and $T_p$ capable of giving life to the spin Seebeck effect *can* be established in longer-wavelength region of the magnon spectrum ($k_m < 10^5$ rad cm$^{-1}$).

In summary, we have made the first measurements of the spatial distribution of the magnetization in a magnetic sample subject to a lateral thermal gradient. In so doing, we have found the lengthscale $\lambda$ which characterizes the magnon-phonon interaction in the magnetic insulator YIG to be significantly smaller than predicted[11]. This result reveals that the coupling of the short wavelength magnons which define the room-temperature magnetization of a magnetic system is far too strong for them to contribute to the spin Seebeck effect (even in YIG, which is known to have significantly weaker magnon-phonon coupling than magnetic metals such as Permalloy in which the experimental observation of this effect was first reported). This suggests that, contrary to what was previously thought, if it exists, the inequality between the magnon and phonon bath temperatures believed to be responsible for the spin Seebeck effect must be *particular* to the longer-wavelength region of the magnon spectrum. Our work not only sheds meaningful new light on the role of thermal magnons in the spin Seebeck effect but potentially lends valuable direction to fundamental and applied spin caloritronics through a new concept of spectrally non-uniform magnon temperature.

# References

1. Bauer, G. E. W., Saitoh, E. & van Wees, B. J. Spin caloritronics. *Nature Mater.* **11,** 391-399 (2012).

2. Uchida, K. *et al.* Observation of the spin Seebeck effect. *Nature* **455,** 778-781 (2008).

3. Costache, M. V., Bridoux, G., Neumann, I. & Valenzuela, S. O. Magnon-drag thermopile. *Nature Mater.* **11,** 199-202 (2012).

4. Uchida, K. *et al.* Long-range spin Seebeck effect and acoustic spin pumping. *Nature Mater.* **10,** 737-741 (2011).

5. Jaworski, C. M. *et al.* Observation of the spin-Seebeck effect in a ferromagnetic semiconductor. *Nature Mater.* **9,** 898-903 (2010).

6. Jaworski, C. M *et al.* Spin-Seebeck effect: a phonon driven spin distribution. *Phys. Rev. Lett.* **106,** 186601 (2011).

7. Uchida, K. *et al.* Spin Seebeck insulator. *Nature Mater.* **9,** 894-897 (2010).

8. Uchida, K., Nonaka, T., Ota, T. & Saitoh, E. Observation of longitudinal spin-Seebeck effect in magnetic insulators. *Appl. Phys. Lett.* **97,** 172505 (2010).

9. Adachi, H. *et al.* Gigantic enhancement of spin Seebeck effect by phonon drag. *Appl. Phys. Lett.* **97,** 252506 (2010).

10. Sanders, D. J. & Walton, D. Effect of magnon-phonon thermal relaxation on heat transport by magnons. *Phys. Rev. B* **15,** 1489-1494 (1977).

11. Xiao, J., Bauer, G. E. W., Uchida, K., Saitoh, E., & Meakawa, S. Theory of magnon-driven spin Seebeck effect. *Phys. Rev. B* **81,** 214418 (2010).

12. Ohe, J., Adachi, H., Takahashi, S. & Maekawa, S. Numerical study on the spin Seebeck effect *Phys. Rev. B* **83,** 115118 (2011).

13. Gurevich, A. G. & Melkov, G. A. *Magnetization Oscillations and Waves* (CRC Press, 1996).

14. Serga, A. A., Chumak, A. V. & Hillebrands, B. YIG magnonics. *J. Phys. D: Appl. Phys.* **43,** 264002 (2010).

15. Sandweg, C. W. *et al.* Wide-range wavevector selectivity of magnon gases in Brillouin light scattering spectroscopy. *Rev. Sci. Instrum.* **81,** 073902 (2010).

16. Doormann, V., Krumme, J. P., Klages, C. P. & Erman, M. Measurement of the refractive index and optical absorption spectra of epitaxial Bismuth substituted Yttrium Iron Garnet films at uv to near-ir wavelengths. *Appl. Phys. A* **34,** 223(1984).

17. Algra, H. A. & Hansen P. Temperature dependence of the saturation magnetization of ion-implanted YIG films. *App. Phys. A* **29**, 83 (1982).

18. Kalinikos, B. A. & Slavin A. N. Theory of dipole-exchange spin wave spectrum for ferromagnetic films with mixed exchange boundary conditions. *J. Phys. C: Solid State Phys.* **19**, 7013 (1986).

19. Kaganov, M. I. & Tsukernik, V. M. Phenomenological theory of kinetic processes in ferromagnetic dielectric. II. Interaction of spin waves with phonons. *Sov. Phys. JETP* **7,** 151 (1959).

**Acknowledgements**

We acknowledge valuable discussions with B. Obry, P. Pirro, A. C. Parra**,** and A. V. Chumak. The research was supported by the Deutsche Forschungsgemeinschaft (SE 1771/4-1) within Priority Program 1538 "Spin Caloric Transport". M.A. is financially supported by the Graduate School of Excellence Materials Science in Mainz through DFG-funding of the Excellence Initiative (GSC 266).

**Author Contributions**

M.A. and V.I.V. designed the experiment, performed the measurements and carried out the data analysis. A.A.S. and B.H. supervised the experiment. A.A.S., B.H. and G.A.M. planned the study and developed the explanation of the results. M.A., V.I.V., A.D.K. and A.A.S. wrote the manuscript. All authors were involved in discussion of the results.

**Additional information**

The authors declare no competing financial interests. Reprints and permissions information is available online at http://www.nature.com/reprints. Correspondence and requests for materials should be addressed to B.H. (hilleb@physik.uni-kl.de)

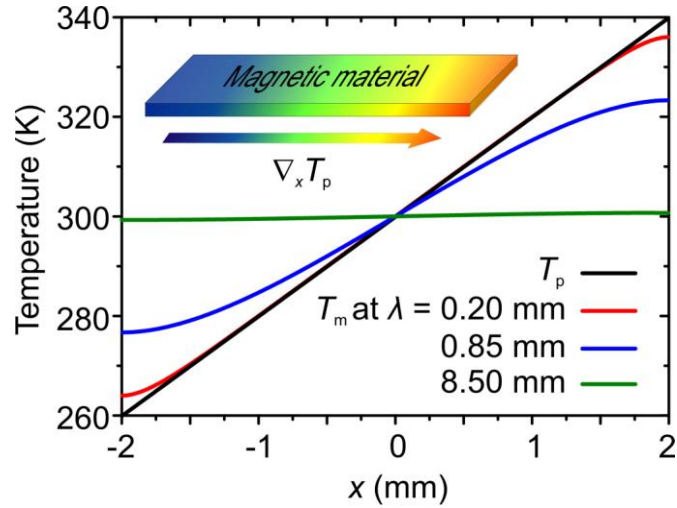

**Figure 1 | Understanding the relationship between magnon and phonon temperatures.** Current understanding of thermoelectric magnetic effects relies on the assumption that magnon-phonon coupling is weak enough to allow the two baths to establish dynamic thermal equilibria at distinct temperatures. In such a situation an applied lateral thermal (phonon) gradient (black line and inset) produces a magnon density gradient in the same direction down which magnons diffuse so that, at equilibrium, the magnon temperature $T_m$ is above the phonon temperature $T_p$ at the cooler end and below it at the hotter one[10,11]. The maximum difference between $T_m$ and $T_p$ depends on the characteristic lengthscale $\lambda$ of the magnon-phonon interaction. Theoretical estimates for $\lambda$ in YIG vary from 0.85 mm to 8.5 mm (coloured lines)[11].

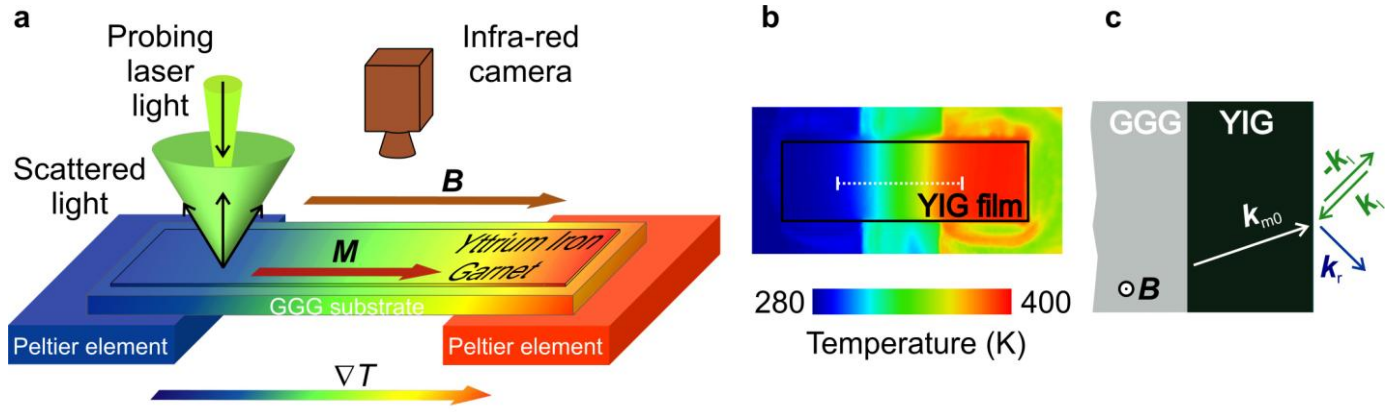

Agrawal *et al.*, Figure 2

**Figure 2 | Magnon temperature measurement setup. a,** In our experiments opposite ends of a YIG film (on a gallium gadolinium garnet substrate) were placed on two Peltier elements 3.2 mm apart to create a lateral thermal gradient $\nabla T$. In the schematic, the coldest regions are shown in blue, the hottest in red. The film was magnetized in-plane with an external magnetic field $B = 250$ mT parallel to $\nabla T$. The magnon temperature was measured using Brillouin light scattering (BLS) spectroscopy. The incident (probing) laser light from the BLS spectrometer and the back scattered (signal) light are represented by the green cones. An infra-red (IR) camera was used to obtain thermal images of the system. **b,** Infra-red image of the YIG film shown in **a**. The white dashed line indicates the BLS measurement path. **c,** The orientation of the incident (probing) ($k_l$), reflected ($k_r$), and back-scattered (signal) (-$k_l$) photon wavevectors relative to the YIG film surface. The magnon wavevector, $k_{m0}$ satisfies the momentum conservation condition, $|k_{m0}| = 2\eta|k_l|$ where $\eta = 2.36$ is the refractive index of the YIG.

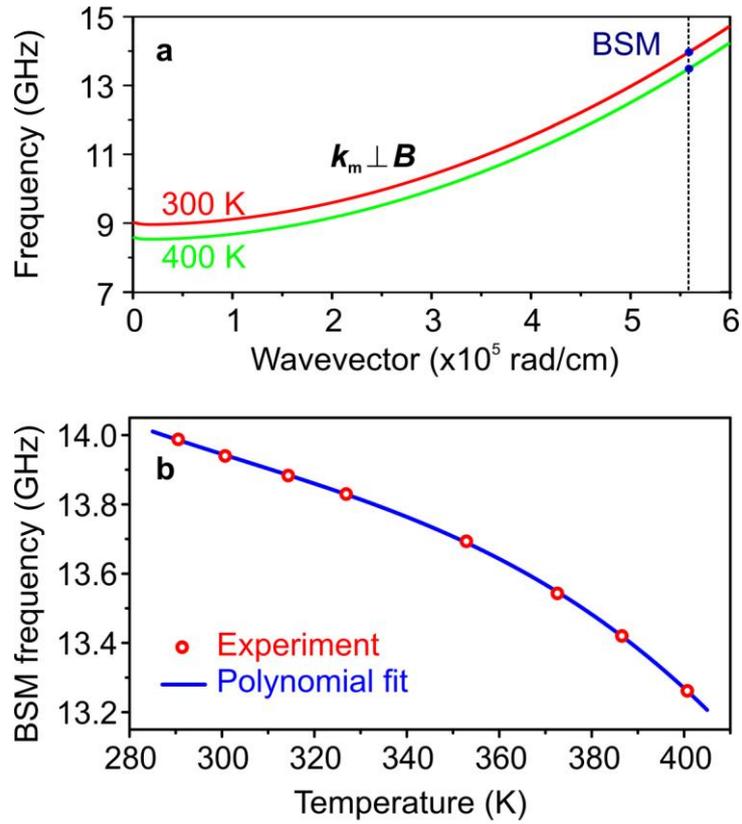

**Figure 3 | Thermal dependence of the back-scattering magnon (BSM) mode. a**, The dispersion relations of magnons propagating perpendicular to the magnetization at temperatures of 300 K and 400 K in a 6.7 μm thick YIG film subject to a bias magnetic field B = 250 mT. Bullets show the positions of the BSM mode at $k_m = 5.67 \times 10^5$ rad cm$^{-1}$. **b**, Measured (open symbols) and polynomially fitted (solid line) BSM frequency as a function of temperature of the YIG film. The frequency decreases monotonically with temperature as a result of the increasing magnon population.

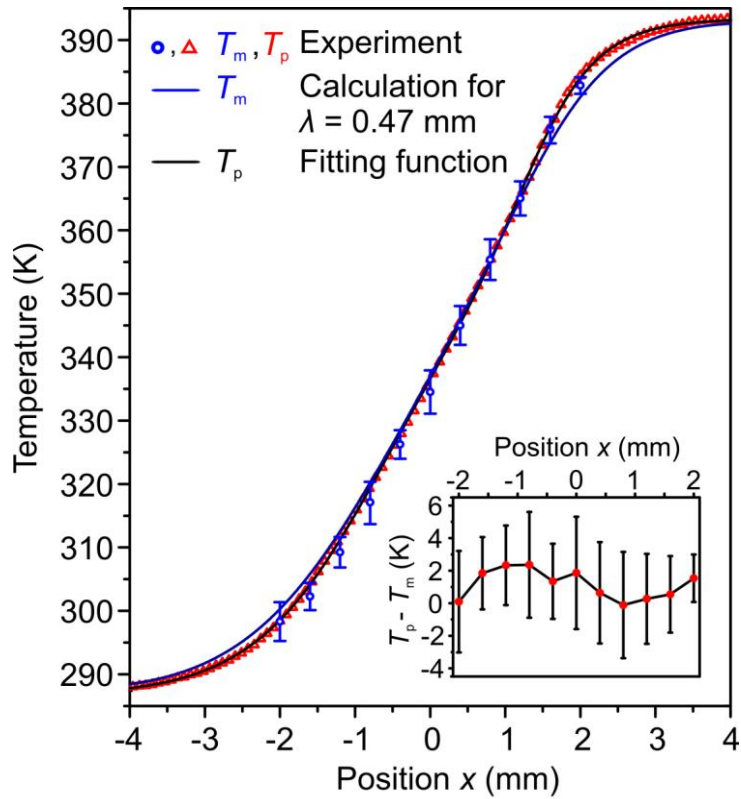

**Figure 4 | Magnon and phonon temperature profiles in a YIG film subject to a thermal gradient.** Plotted are the measured phonon and magnon temperatures $T_p$ (open triangles) and $T_m$ (open circles) along the BLS laser scan line indicated in Fig. 2(b). The $T_p$ data is fitted with a Boltzmann sigmoid function. The profile of $T_m(x)$ is numerically calculated for the characteristic length parameter $\lambda = 0.47$ mm. The inset shows the difference between $T_p$ and $T_m$ with a 95% of confidence level.